# Ultra-Low-Power Tuning in Hybrid Barium Titanate−Silicon Nitride Electro-optic Devices on Silicon

J. Elliott Ortmann,*,†,‡ Felix Eltes,‡ Daniele Caimi,‡ Norbert Meier,‡ Alexander A. Demkov,† Lukas Czornomaz,‡ Jean Fompeyrine,‡ and Stefan Abel*,‡

†Department of Physics, The University of Texas, Austin, Texas 78712, United States
‡IBM Research−Zurich, Säumerstrasse 4, 8803 Rüschlikon, Switzerland

*Supporting Information*

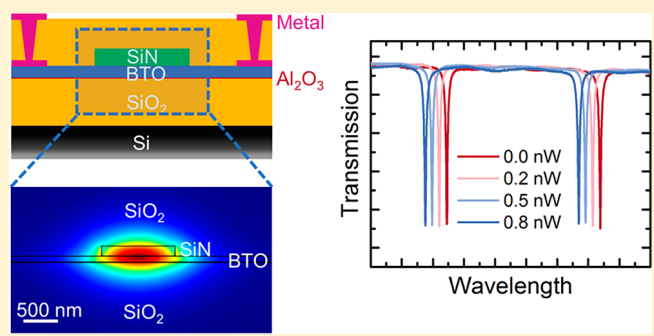

**ABSTRACT:** As the optical analogue to integrated electronics, integrated photonics has already found widespread use in data centers in the form of optical interconnects. As global network traffic continues its rapid expansion, the power consumption of such circuits becomes a critical consideration. Electrically tunable devices in photonic integrated circuits contribute significantly to the total power budget, as they traditionally rely on inherently power-consuming phenomena such as the plasma dispersion effect or the thermo-optic effect for operation. Here, we demonstrate ultra-low-power refractive index tuning in a hybrid barium titanate (BTO)−silicon nitride (SiN) platform integrated on silicon. We achieve tuning by exploiting the large electric field-driven Pockels effect in ferroelectric BTO thin films of sub-100 nm thickness. The extrapolated power consumption for tuning a free spectral range (FSR) in racetrack resonator devices is only 106 nW/FSR, several orders of magnitude less than many previous reports. We demonstrate the technological potential of our hybrid BTO−SiN technology by compensating thermally induced refractive index variations over a temperature range of 20 °C and by using our platform to fabricate tunable multiresonator optical filters. Our hybrid BTO−SiN technology significantly advances the field of ultra-low-power integrated photonic devices and allows for the realization of next-generation efficient photonic circuits for use in a variety of fields, including communications, sensing, and computing.

Decades ago, the development of integrated electronic circuits revolutionized the electronics industry by enabling electronic circuits to be made smaller, faster, and more cheaply. A similar trend has reached optical networks, where the integration of bulk- and fiber-based circuits can nowadays be realized in photonic integrated circuits (PICs) on compact semiconductor substrates. PICs already play a major role in data centers, and their use is expected to rapidly expand in the coming years owing to the explosive growth of global network traffic.[1−3] In addition to well-established transceiver technologies for data centers, PICs also show promise for use in a wide variety of emerging fields, such as reservoir computing,[4] deep learning,[5] photonic quantum computing,[6] and sensing technologies.[7−9]

While many materials have been explored for PICs, silicon-[10,11] and silicon nitride (SiN)-based[12,13] PIC technologies have received widespread attention due to their compatibility with existing complementary metal-oxide-semiconductor (CMOS) processes. CMOS compatibility is a critical consideration for technological adoption, as it allows for the reliable and cost-efficient construction of micro- and nanoscale devices and the co-integration of electronic and photonic circuitry on a single chip. Besides CMOS compatibility, both silicon and SiN have their own unique advantages when employed in PICs. For example, silicon offers excellent light confinement and a variety of nonlinear effects that can be useful for the construction of a multitude of photonic devices.[11] On the other hand, SiN can accommodate a broad wavelength range, including visible light, can be stacked to form multiplanar device architectures,[14] and can be engineered for very low optical propagation losses.[15,16]

Regardless of one's material platform of choice, a key functionality needed in PICs is the ability to tune the effective refractive index of the guided optical signal. Such refractive index tuning is the building block for electro-optic modulators,[17] switches,[18] and filters,[19−22] which are all fundamental components of PICs. Additionally, refractive index tuning can be used to compensate for unavoidable fabrication imperfections[23−26] and thermal fluctuations[27,28] that erode device performance. Such compensation is particularly important for resonant devices, which offer very small footprints but also suffer from a high sensitivity to environmental changes and fabrication tolerances.

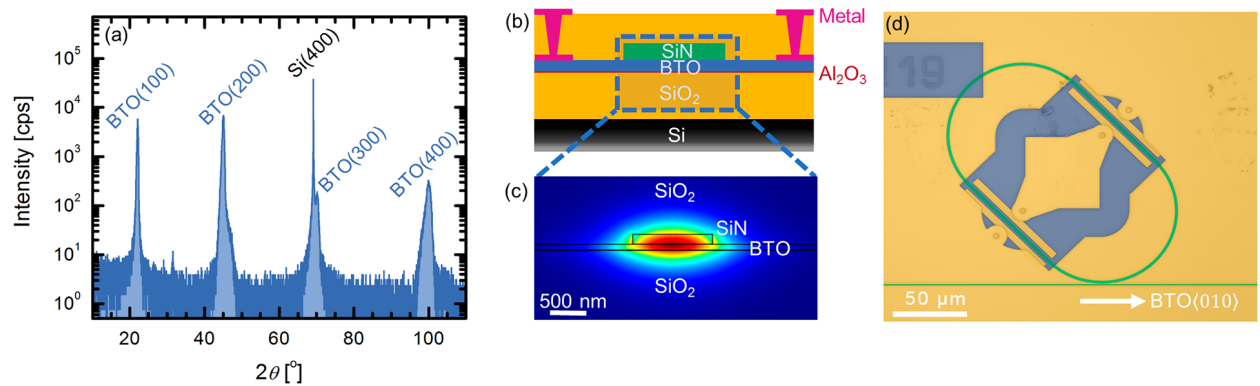

**Figure 1.** (a) Out-of-plane $\theta/2\theta$ X-ray diffraction scan showing an epitaxial BTO thin film on silicon. (b) Cross-sectional schematic of our hybrid BTO–SiN devices. The BTO layer is 80 nm thick, while the SiN waveguide is 150 nm thick and 1.1 $\mu$m wide. (c) Simulated fundamental TE mode in a hybrid BTO–SiN waveguide. The color mapping indicates the norm of the electric field, where red represents regions of high field and blue represents regions of low field. (d) False-color optical micrograph of a hybrid BTO–SiN racetrack resonator used for index tuning experiments, showing the SiN strip waveguides (green), metal electrodes (yellow), and regions of $SiO_2$ cladding not covered with metal (blue). The white arrow indicates the crystalline orientation of the BTO layer.

Several different mechanisms have been exploited for tuning the effective refractive index in integrated photonic devices. The most prominent are the thermo-optic effect, which is present in both silicon- and SiN-based platforms,[28−30] and the plasma dispersion effect, which is available in silicon-based devices.[17,31] Unfortunately, both tuning mechanisms are inherently power consuming. Thermo-optic tuning relies on Joule heating and therefore necessitates significant current flow. Similarly, forward-biased plasma dispersion silicon photonic devices require current flow through a p–n junction. State-of-the-art resonator devices relying on the thermo-optic and plasma dispersion effects for tuning show power consumption on the order of 0.1–100 mW/free spectral range (FSR).[27−29] The static tuning power of such devices can be several times larger than their dynamic power consumption,[32−34] placing severe restrictions on the use of resonant devices in a variety of applications including high-throughput data centers and next-generation sensing technology. Reverse-biased devices are generally less power consuming than forward-biased devices, falling on the lower end of the power consumption scale cited above,[27] but can induce large optical propagation losses caused by high doping levels and are generally less suited for tuning due to the small accessible refractive index window. By introducing III–V layers and forming hybrid III–V/silicon phase shifters, these loss issues have recently been significantly improved.[35] However, the delicate process of wafer-bonding III–V layers to silicon waveguides complicates fabrication, while the unavailability of 200 mm InP wafers limits the scalability of the hybrid III–V/silicon devices. Furthermore, the transparency window of hybrid III–V/silicon devices is limited to wavelengths greater than ∼1.3 $\mu$m by the III–V layers, in contrast to the large transparent window available in hybrid BTO–SiN devices, which should extend into the visible spectrum.

In contrast to the thermo-optic and plasma dispersion effects, the Pockels effect is a second-order nonlinear optic effect describing the electric field-driven modulation of the refractive index. The Pockels effect is therefore uniquely suited for use in ultra-low-power integrated photonics applications, as no electric current flow or doped regions in the waveguide are required for operation. As neither silicon nor SiN inherently displays a strong Pockels effect,[36−38] other materials must be combined with the waveguides. The ferroelectric perovskite $BaTiO_3$ (barium titanate, BTO) is an excellent candidate, as it displays a very large Pockels response even in thin-film form[39] and can be readily integrated with silicon via an epitaxial $SrTiO_3$ buffer layer.[40,41] Furthermore, low-loss hybrid BTO–silicon waveguides have already been demonstrated,[42] as well as electro-optic operation in hybrid BTO–silicon structures[39,43,44] on 200 mm wafer sizes.[45,46] Other platforms, such as those based on the transition metal oxides lead zirconate titanate[47] and lithium niobate,[48] have also been explored for integrated Pockels devices. However, their performance in regard to static power consumption is still unclear.

While hybrid BTO–silicon photonic devices have recently garnered significant attention, the combination of BTO with SiN potentially offers many advantages over a silicon-based platform. For example, the integration of SiN strip waveguides with BTO allows for the combination of ultra-low-power refractive index tuning via the Pockels effect in BTO with the low optical losses available in SiN.[15,16,49] In addition, there are no mobile charge carriers in highly insulating SiN that can impact electro-optic performance. Finally, due to the different material absorption, the optical wavelength range available in BTO–SiN waveguides is significantly wider than in BTO–Si waveguides, allowing operation in the visible wavelength range, for example. Electro-optic devices featuring BTO integrated on MgO have previously been demonstrated at a wavelength of 632 nm, indicating that BTO maintains a strong electro-optic response in the visible spectrum.[50−53] In this work, we fabricate hybrid BTO–SiN racetrack resonators and demonstrate the ability to electrically tune the effective refractive index on the order of $10^{-3}$. Our devices feature a remarkably low power consumption of approximately 106 nW/FSR, several orders of magnitude lower than many previous reports.[27−29,44,54] We demonstrate the technological potential of our hybrid BTO–SiN technology by compensating thermal refractive index variations over a temperature range of 20 °C as well as by demonstrating tunable multiresonator optical filters that can be used to compensate for unavoidable fabrication imperfections.

## ■ METHODS

**Device Fabrication.** We epitaxially grow 80 nm single-crystalline BTO thin films (Figure 1a) by molecular beam epitaxy on silicon-on-insulator substrates and subsequently transfer them to thermally oxidized silicon wafers via wafer-bonding.[39] The device silicon layer of the donor wafer is removed by dry chemical etching. Next, 150 nm SiN is deposited on the BTO layer by plasma-enhanced chemical vapor deposition (PECVD) and patterned to form waveguides using e-beam lithography and reactive ion etching. Finally, a combination of tungsten deposition, $SiO_2$ deposition, and dry chemical etching processes were used to form side electrodes, vias, cladding, and metal pads for contacting the devices, resulting in a final device structure shown schematically in Figure 1b.

**Electro-optic Characterization.** Light from a fiber-coupled, tunable continuous-wave laser operating around 1.55 $\mu$m was coupled into and out of the devices via grating couplers and detected with a power meter. The incoming polarization was set using a polarization controller. A parameter analyzer was used to apply the bias to the devices and to determine the leakage currents. Device temperature was adjusted and monitored via an external temperature controller for temperature-dependent measurements. A detailed schematic of the setup used for electro-optic characterization is presented in the Supporting Information Section 1.

We measured the resonance shifts $\Delta\lambda$ for different electric fields and converted them to effective refractive index shifts $\Delta n_{eff}$ using the relation $\Delta n_{eff} = \lambda_0 \Delta\lambda/\mathrm{FSR}(L_e)$, where $\lambda_0$ is the resonance wavelength at 0 V bias, FSR is the free-spectral range in terms of wavelength, and $L_e$ is the total length of the electrodes.

**Optical Mode Simulations.** Optical mode simulations were performed using COMSOL Multiphysics and the COMSOL RF module. The refractive indices used for the simulations are presented in Supporting Information Table S1. The perfect electrical conductor boundary condition (i.e., $\vec{n} \times \vec{E} = 0$, where $\vec{n}$ is the unit normal vector and $\vec{E}$ is the electric field vector) was applied to all outer boundaries of the 2d simulation cell. The electric field was taken to be perfectly in-plane initially, and the wavelength was taken to be 1550 nm.

## ■ RESULTS AND DISCUSSION

**Ultra-Low-Power Refractive Index Tuning.** The waveguides of the fabricated devices support a single TE-like mode (Figure 1c) with 18% of the optical power confined in the electro-optically active BTO layer. Power confinement in the BTO scales with BTO thickness and can therefore be increased by fabricating devices with thicker BTO layers (Supporting Information Section 2). The propagation losses are 9.4 dB/cm (Supporting Information Section 3), comparable to previous reports of hybrid BTO–silicon waveguides,[39,42,45] and can likely be reduced through the development of an optimized fabrication process.[16,42] Racetrack resonators used for index tuning experiments feature two 75-$\mu$m-long straight segments with electrodes separated by 5.1 $\mu$m to apply an electric field for electro-optic operation (Figure 1d). We designed the electrodes such that the electric field is applied along the BTO$\langle 011 \rangle$ family of lattice vectors in order to exploit the largest Pockels coefficient, $r_{42}$, and maximize the electro-optic response.[39]

The transmission spectra of the devices (Figure 2a) feature sharp resonances with extinction ratios of approximately 18 dB.

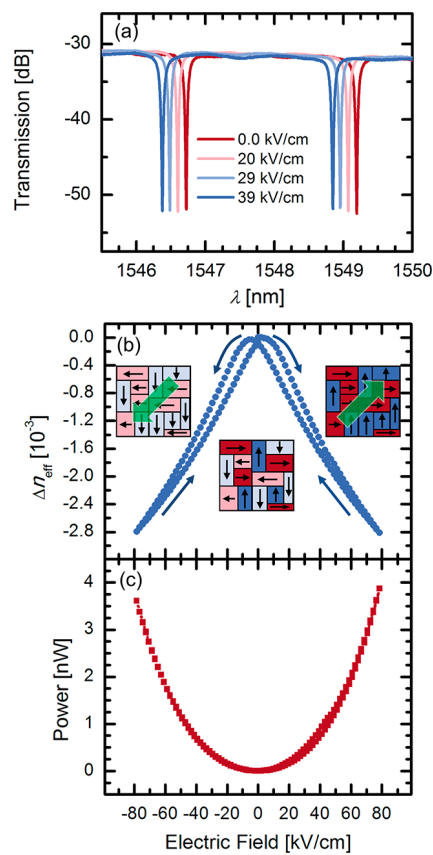

Figure 2. (a) Transmission spectra of a racetrack resonator for different electric fields. The position of the resonance is shifted due to the Pockels effect in BTO. (b) Change of the effective refractive index in the active phase shifter section as a function of applied electric field. As expected, the ferroelectric nature of BTO results in a hysteresis in the electro-optic response. Insets are schematics of the top view of the ferroelectric domain structure in the poled (left and right) and unpoled (center) states. The BTO films show four types of ferroelectric domains with the ferroelectric polarization (black arrow) corresponding to the vertical (blue) and horizontal (red) directions.[39,41] Green arrows represent the applied electric field. (c) Power consumption as a function of the applied field recorded during measurement of the hysteresis loop shown in (b). The power function is defined as the leakage current times the applied voltage.

Upon applying an electric field across the device, the resonance positions shift, indicating a change in the effective refractive index within the resonator. The electric field $E$ across the waveguide was estimated according to $E = V/d$, where $V$ is the applied bias and $d$ is the separation between electrodes. Following the procedure outlined in the Methods section, we track the shift of the effective index as a function of applied field (Figure 2b). The refractive index varies linearly with the applied field above fields of approximately 40 kV/cm, as expected for the Pockels effect. At lower fields, the refractive index variation deviates from the linear behavior due to nonlinear contributions from ferroelectric domain switching.[39,41,45] Such domain switching results in hysteretic behavior of the refractive index when sweeping the electric field (Figure 2b), in line with previous reports.[39,41,45]

An effective Pockels coefficient of approximately 343 pm/V is extracted from the index tuning experiments, which agrees with previous reports on BTO thin films.[46] Due to the strong dependence of the effective Pockels coefficient on the crystalline quality, morphology, and symmetry, which in turn strongly depend on deposition conditions[55] and film thicknesses,[56,57] both larger[39] and smaller[41,43,55] Pockels coefficients in a similar range to the value determined in this study have been reported in the past.

The application of a bias across the device does not alter the extinction ratios even up to 40 V (78 kV/cm). This behavior is expected, as the Pockels effect does not impact the imaginary part of the refractive index. In contrast, active operation of devices exploiting the plasma dispersion effect typically results in undesired, additional optical absorption.[27,58] Our devices show ultra-low power consumption (Figure 2c) of 106 ± 5 nW/FSR (Supporting Information Section 4), several orders of magnitude less than that in plasma dispersion- and thermo-optic-based electro-optic tuning devices that are compatible with standard PIC integration routes (Table 1).[27−29,54]

Table 1. Power Consumption Comparison between Electro-optic Devices Compatible with Standard PIC Integration Routes, Exploiting Different Tuning Mechanisms

| tuning mechanism | reference | power consumption ($\mu$W/FSR) |
|---|---|---|
| thermo-optic | Atabaki et al.[29] | 24 500 |
| thermo-optic | Dong et al.[54] | 21 000 |
| thermo-optic | Dong et al.[28] | 2400 |
| plasma dispersion in Si[a] | Timurdogan et al.[27] | >500 |
| Pockels effect | Abel et al.[44] | 4 |
| Pockels effect | this work | 0.106 |

[a]Power consumption extrapolated from tuning data presented in Timurdogan et al.

Furthermore, analogous electro-optic tuning devices utilizing the Pockels effect in lithium niobate integrated with silicon and SiN strip waveguides have been demonstrated, although with significantly smaller tunability than the devices reported here.[59−61] While the power consumption of the lithium niobate devices is not explicitly stated, the large voltages required for tuning suggest greater static power consumption than the BTO devices reported here.

It should be noted that a full FSR of tuning is not achieved in the devices reported here. However, the tuning range can be significantly increased for a given electric field and the resonator devices can be optimized for low-bias operation. Example optimizations include improvements to the electrode design, such as the use of transparent conducting electrodes,[62] and the use of thicker BTO in future devices, thereby increasing the optical confinement in BTO as well as the electro-optic tunability (Supporting Information Section 2). We estimate the power consumption of the optimized resonators to be approximately 22 nW/FSR. Furthermore, the BTO−SiN platform can be used to construct linear phase shifters that are well-suited for low-bias operation due to their much greater active length relative to compact resonator devices. Our calculations suggest $V_\pi$ = 3 V for a 1-mm-long linear phase shifter constructed from the BTO−SiN platform with a corresponding power consumption $P_\pi \approx$ 1 nW (Supporting Information Section 2). A similar power consumption was recently reported for hybrid III−V/silicon linear phase shifters.[35] However, the small band gap of the III−V materials and challenging fabrication route significantly limit the optically transparent window and integration flexibility relative to the hybrid BTO−SiN platform discussed in this work (Table 1).

The measured power consumption is the result of a small leakage current across the nominally insulating BTO layer (Supporting Information Section 5). The leakage current in oxides is typically associated with a Pool−Frenkel-type conduction mechanism.[63] The origin of the charge transport are defects in the material such as oxygen vacancies or interfacial traps, which are determined by the material deposition and device fabrication process. Indeed, oxides that can be used for Pockels-active devices can show significant power consumption due to relatively high leakage currents.[44,64] The realization of exceptionally low leakage current on the order of pA even after device processing and PECVD of SiN is a major achievement and indicates the great promise of the hybrid BTO−SiN platform in realistic, power-sensitive technologies.

**Thermal Compensation.** Having established ultra-low-power refractive index tuning in the hybrid BTO−SiN electro-optic devices, we use the platform to demonstrate two example applications of this technology. First, we use racetrack resonator devices to compensate for thermal refractive index variations. A key challenge facing future integrated photonics technology is the ability to sustain reliable operation in environments subject to temperature fluctuations.[65] Because SiN[30] and in particular silicon[66−68] show a significant thermo-optic effect, relatively minor temperature fluctuations can have a significant impact on the effective refractive index of guided modes. In resonators, thermally induced index fluctuations can therefore alter a device's optical characteristics at a given wavelength unless such index fluctuations are appropriately compensated. Ultra-low-power Pockels tuning of the effective refractive index allows for an efficient way to compensate for thermal index drift without dissipating additional heat across the device, as would occur in devices exploiting current-driven tuning mechanisms.

To demonstrate thermal refractive index drift compensation via Pockels tuning, we analyzed the transmission of the devices in the temperature range from room temperature (25 °C) up to 45 °C (Figure 3a). We extracted a thermo-optic resonance shift in our devices of 21.6 pm/°C. At each stable temperature point, we tuned the bias appropriately such as to shift the resonance back to its room-temperature position (Figure 3b) and measured the power dissipated across the device at the optimal tuning voltage. Through this procedure, we compensate for thermal refractive index variation over a temperature range of 20 °C while consuming less than 1 nW of static power, 4 orders of magnitude less than the approximately 50 $\mu$W of static power required to compensate for heating of 10 °C in state-of-the-art p−n modulators.[27] Importantly, the minimal power dissipated across our devices during electro-optic operation results in negligible heating, allowing for the compensation of large temperature fluctuations without introducing additional heating from device operation. The negligible heating in our devices also reduces thermal cross-talk as compared to thermo-optic devices,[69] allowing for exceedingly local refractive index tuning even for dense device arrays. The relatively high Curie temperature of BTO allows for high-temperature operation in BTO-based electro-optic devices, with the unimpaired operation of BTO-on-silicon plasmonic modulators having been demonstrated up to 130 °C.[70]

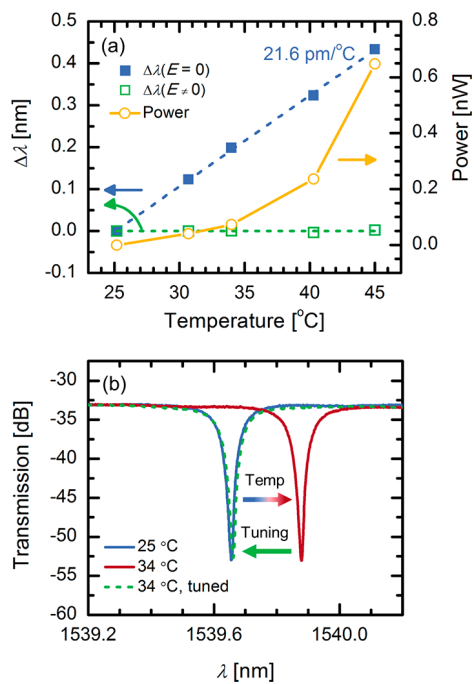

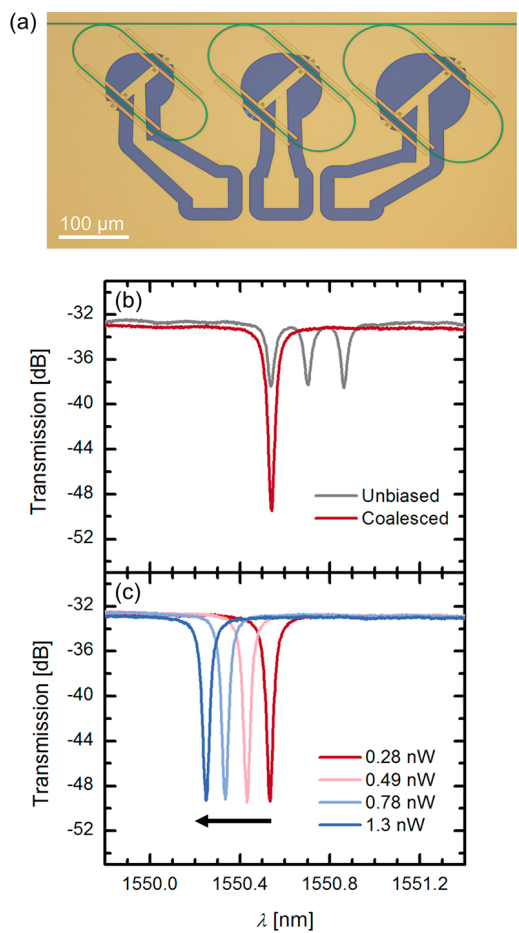

Figure 3. (a) Zero-field (blue) and electrically tuned (green) resonance wavelength shift and static tuning power (yellow) as a function of temperature. Dashed lines are linear fits to the data. (b) Racetrack resonator spectra showing the room-temperature resonance (blue) and the same resonance at 34 °C, both before (red, solid line) and after (green, dashed line) tuning.

**Tunable Multiresonator Filters.** As a second example application of our hybrid BTO–SiN platform, we fabricated coupled multiresonator optical filters. Many photonics applications, including quantum information processing,[20] wavelength division multiplexing,[19] and on-chip reconfigurable networks[22] make use of high-quality optical filters. While many reports have explored the use of coupled resonators as optical filters,[71–74] resonators suffer from an extreme sensitivity to fabrication tolerances, and even minor imperfections can decouple resonances.[23–26] However, by incorporating an optical path length trimming mechanism into the devices to compensate for fabrication imperfections, coupled resonator filters can be made significantly more robust.

We have utilized our BTO-SiN platform to fabricate tunable, multiracetrack resonator optical filters (Figure 4a). The devices feature three racetrack resonators coupled to a straight waveguide, although the design can be extended to a larger number of resonators. Because the purpose of these experiments is to demonstrate the ability to overcome fabrication imperfections in multiresonator filters via low-power electro-optic tuning, we focused on cascaded multiresonator filters, rather than coupled resonator filters,[71] for simplicity. Differences in resonator cavity lengths result in the appearance of three distinct resonances (Figure 4b), one for each resonator. By appropriately biasing two of the resonators, the individual resonances can be coalesced into a single resonance whose extinction ratio is equal to the sum of the extinction ratios of the individual resonances (Figure 4b). Here, we have purposely designed devices in which the extinction ratios of the individual resonances are only 6 dB to ensure the minimum transmission within the coalesced resonance is still measurable in our experimental setup. However, the multiresonator tunable filters presented here can in principle be used to fabricate devices with very large extinction ratios. The coalesced resonance can itself also be spectrally tuned (Figure 4c) by appropriately biasing all three resonators. The linear tuning behavior of the refractive index with the electric field (Figure 2b) simplifies the determination of the individual resonator bias conditions required for tuning the coalesced resonance.

Figure 4. (a) False-color optical micrograph of a multiresonator tunable filter, showing the SiN strip waveguides (green), metal electrodes (yellow), and regions of SiO$_2$ cladding not covered with metal (blue). (b) Unbiased (gray) and biased (red) spectra of the multiresonator filter, demonstrating the ability to coalesce individual resonances into a single resonance. (c) Coalesced resonance shifted by an external electric field. The arrow indicates the tuning direction, and the legend indicates the total tuning power.

## ■ CONCLUSION

In summary, we have demonstrated ultra-low-power refractive index tuning in silicon-integrated hybrid BTO–SiN photonic devices. By exploiting the large electric field-driven Pockels effect in BTO, we have realized index tuning with static power consumption of approximately 106 nW/FSR in racetrack resonator devices. We use our devices to compensate for thermal refractive index variations and demonstrate the ability to compensate for heating of 20 °C while consuming less than 1 nW of static power. Additionally, we fabricate multiresonator optical filters and provide a proof-of-concept demonstration for using the Pockels effect to perform optical path length trimming and overcome decoupling arising from fabrication imperfections.

The devices we present outperform previously reported tuning elements in terms of power consumption by several orders of magnitude. Furthermore, the combination of BTO with SiN enables fully dielectric waveguides with very low propagation losses that extend the optically transparent window into the visible range, allowing the hybrid BTO–SiN platform to find use in a wide variety of next-generation technologies including biosensors[75] and LIDAR.[76,77] Our results show the technological potential of the Pockels-active hybrid BTO–SiN platform for ultra-low-power integrated photonic devices. The availability of such devices in integrated photonics shines new light onto the implementation of resonant structures in compact PICs. While such structures offer a variety of attractive properties, such as small footprints and sharp transmission spectra as needed, for example, for filters, their applicability has always been limited by the power consumption needed for active tuning. Hybrid BTO–SiN devices can overcome this issue and enable the next generation of photonic applications, including compact communications technologies, optical computing systems, and ultrasensitive photonic sensors.

## ■ ASSOCIATED CONTENT

**ⓈSupporting Information**

The Supporting Information is available free of charge on the ACS Publications website at DOI: 10.1021/acsphotonics.9b00558.

> Experimental details of electro-optical characterization, optical mode simulations, additional electrical, optical, and electro-optical characterization of devices, and calculations describing opportunities for device performance optimization (PDF)

## ■ AUTHOR INFORMATION

**Corresponding Authors**
*E-mail: jortmann13@gmail.com.
*E-mail: sab@zurich.ibm.com.

**ORCID**
J. Elliott Ortmann: 0000-0001-6603-1238
Felix Eltes: 0000-0003-3882-8121
Alexander A. Demkov: 0000-0003-4241-3519

**Notes**
The authors declare no competing financial interest.

## ■ ACKNOWLEDGMENTS

J.E.O. is grateful for the generous support of the National Science Foundation Graduate Research Fellowship under grant no. DGE-1610403. J.E.O. and A.A.D. gratefully acknowledge support from the National Science Foundation under grant no. IRES-1358111 and the Air Force Office of Scientific Research under grants FA9550-12-10494 and FA9550-18-1-0053. This project has received funding from the European Commission under grant agreements H2020-ICT-2015-25-688579 (PHRESCO) and H2020-ICT-2017-1-780997 (plaCMOS), from the Swiss State Secretariat for Education, Research and Innovation under contract no. 15.0285, and from the Swiss National Foundation project no. 200021_159565 PADOMO.

## ■ REFERENCES

(1) Lam, C.; Liu, H.; Koley, B.; Zhao, X.; Kamalov, V.; Gill, V. Fiber Optic Communication Technologies: What's Needed for Datacenter Network Operations. *IEEE Commun. Mag.* **2010**, *48*, 32−39.
(2) Thomson, D.; Zilkie, A.; Bowers, J. E.; Komljenovic, T.; Reed, G. T.; Vivien, L.; Marris-Morini, D.; Cassan, E.; Virot, L.; Fédéli, J.-M.; et al. Roadmap on Silicon Photonics. *J. Opt.* **2016**, *18*, 073003.
(3) Zhou, Z.; Chen, R.; Li, X.; Li, T. Development Trends in Silicon Photonics for Data Centers. *Opt. Fiber Technol.* **2018**, *44*, 13−23.
(4) Vandoorne, K.; Mechet, P.; Van Vaerenbergh, T.; Fiers, M.; Morthier, G.; Verstraeten, D.; Schrauwen, B.; Dambre, J.; Bienstman, P. Experimental Demonstration of Reservoir Computing on a Silicon Photonics Chip. *Nat. Commun.* **2014**, *5*, 3541.
(5) Shen, Y.; Harris, N. C.; Skirlo, S.; Prabhu, M.; Baehr-Jones, T.; Hochberg, M.; Sun, X.; Zhao, S.; Larochelle, H.; Englund, D.; et al. Deep Learning with Coherent Nanophotonic Circuits. *Nat. Photonics* **2017**, *11*, 441−446.
(6) Qiang, X.; Zhou, X.; Wang, J. B. J.; Wilkes, C. M.; Loke, T.; O'Gara, S.; Kling, L.; Marshall, G. D.; Santagati, R.; Ralph, T. C.; et al. Large-Scale Silicon Quantum Photonics Implementing Arbitrary Two-Qubit Processing. *Nat. Photonics* **2018**, *12*, 534−539.
(7) Leinders, S. M.; Westerveld, W. J.; Pozo, J.; van Neer, P. L. M. J.; Snyder, B.; O'Brien, P.; Urbach, H. P.; de Jong, N.; Verweij, M. D. A Sensitive Optical Micro-Machined Ultrasound Sensor (OMUS) Based on a Silicon Photonic Ring Resonator on an Acoustical Membrane. *Sci. Rep.* **2015**, *5*, 14328.
(8) Dutt, A.; Joshi, C.; Ji, X.; Cardenas, J.; Okawachi, Y.; Luke, K.; Gaeta, A. L.; Lipson, M. On-Chip Dual-Comb Source for Spectroscopy. *Sci. Adv.* **2018**, *4*, e1701858.
(9) Kita, D. M.; Miranda, B.; Favela, D.; Bono, D.; Michon, J.; Lin, H.; Gu, T.; Hu, J. High-Performance and Scalable on-Chip Digital Fourier Transform Spectroscopy. *Nat. Commun.* **2018**, *9*, 4405.
(10) Hochberg, M.; Baehr-Jones, T. Towards Fabless Silicon Photonics. *Nat. Photonics* **2010**, *4*, 492−494.
(11) Leuthold, J.; Koos, C.; Freude, W. Nonlinear Silicon Photonics. *Nat. Photonics* **2010**, *4*, 535−544.
(12) Moss, D. J.; Morandotti, R.; Gaeta, A. L.; Lipson, M. New CMOS-Compatible Platforms Based on Silicon Nitride and Hydex for Nonlinear Optics. *Nat. Photonics* **2013**, *7*, 597−607.
(13) Muñoz, P.; Micó, G.; Bru, L.; Pastor, D.; Pérez, D.; Doménech, J.; Fernández, J.; Baños, R.; Gargallo, B.; Alemany, R.; et al. Silicon Nitride Photonic Integration Platforms for Visible, Near-Infrared and Mid-Infrared Applications. *Sensors* **2017**, *17*, 2088.
(14) Chiles, J.; Buckley, S. M.; Nam, S. W.; Mirin, R. P.; Shainline, J. M. Design, Fabrication, and Metrology of 10 × 100 Multi-Planar Integrated Photonic Routing Manifolds for Neural Networks. *APL Photonics* **2018**, *3*, 106101.
(15) Stutius, W.; Streifer, W. Silicon Nitride Films on Silicon for Optical Waveguides. *Appl. Opt.* **1977**, *16*, 3218.
(16) Domínguez Bucio, T.; Khokhar, A. Z.; Lacava, C.; Stankovic, S.; Mashanovich, G. Z.; Petropoulos, P.; Gardes, F. Y. Material and Optical Properties of Low-Temperature $NH_3$-Free PECVD $SiN_x$ Layers for Photonic Applications. *J. Phys. D: Appl. Phys.* **2017**, *50*, 025106.
(17) Reed, G. T.; Mashanovich, G.; Gardes, F. Y.; Thomson, D. J. Silicon Optical Modulators. *Nat. Photonics* **2010**, *4*, 518−526.
(18) Nikolova, D.; Rumley, S.; Calhoun, D.; Li, Q.; Hendry, R.; Samadi, P.; Bergman, K. Scaling Silicon Photonic Switch Fabrics for Data Center Interconnection Networks. *Opt. Express* **2015**, *23*, 1159.
(19) Horst, F.; Green, W. M.; Assefa, S.; Shank, S. M.; Vlasov, Y. A.; Offrein, B. J. Cascaded Mach-Zehnder Wavelength Filters in Silicon Photonics for Low Loss and Flat Pass-Band WDM (de-)Multiplexing. *Opt. Express* **2013**, *21*, 11652.
(20) Harris, N. C.; Grassani, D.; Simbula, A.; Pant, M.; Galli, M.; Baehr-Jones, T.; Hochberg, M.; Englund, D.; Bajoni, D.; Galland, C. Integrated Source of Spectrally Filtered Correlated Photons for Large-Scale Quantum Photonic Systems. *Phys. Rev. X* **2014**, *4*, 041047.

(21) Zheng, S.; Zhou, N.; Long, Y.; Ruan, Z.; Du, J.; Hu, X.; Shen, L.; Li, S.; Wang, J. Compact Tunable Photonic Comb Filter on a Silicon Platform. *Opt. Lett.* 2017, 42, 2762.

(22) Wang, S.; Feng, X.; Gao, S.; Shi, Y.; Dai, T.; Yu, H.; Tsang, H.-K.; Dai, D. On-Chip Reconfigurable Optical Add-Drop Multiplexer for Hybrid Wavelength/Mode-Division-Multiplexing Systems. *Opt. Lett.* 2017, 42, 2802.

(23) Prinzen, A.; Waldow, M.; Kurz, H. Fabrication Tolerances of SOI Based Directional Couplers and Ring Resonators. *Opt. Express* 2013, 21, 17212.

(24) Pathak, S.; Van Thourhout, D.; Bogaerts, W. Design Trade-Offs for Silicon-on-Insulator-Based AWGs for (de)Multiplexer Applications. *Opt. Lett.* 2013, 38, 2961.

(25) Dwivedi, S.; D'heer, H.; Bogaerts, W. Maximizing Fabrication and Thermal Tolerances of All-Silicon FIR Wavelength Filters. *IEEE Photonics Technol. Lett.* 2015, 27, 871−874.

(26) Bogaerts, W.; Chrostowski, L. Silicon Photonics Circuit Design: Methods, Tools and Challenges. *Laser Photon. Rev.* 2018, 12, 1700237.

(27) Timurdogan, E.; Sorace-Agaskar, C. M.; Sun, J.; Shah Hosseini, E.; Biberman, A.; Watts, M. R. An Ultralow Power Athermal Silicon Modulator. *Nat. Commun.* 2014, 5, 4008.

(28) Dong, P.; Qian, W.; Liang, H.; Shafiiha, R.; Feng, D.; Li, G.; Cunningham, J. E.; Krishnamoorthy, A. V.; Asghari, M. Thermally Tunable Silicon Racetrack Resonators with Ultralow Tuning Power. *Opt. Express* 2010, 18, 20298.

(29) Atabaki, A. H.; Eftekhar, A. A.; Yegnanarayanan, S.; Adibi, A. Sub-100-ns Thermal Reconfiguration of Silicon Photonic Devices. *Opt. Express* 2013, 21, 15706.

(30) Arbabi, A.; Goddard, L. L. Measurements of the Refractive Indices and Thermo-Optic Coefficients of $Si_3N_4$ and $SiO_x$ Using Microring Resonances. *Opt. Lett.* 2013, 38, 3878.

(31) Soref, R.; Bennett, B. Electrooptical Effects in Silicon. *IEEE J. Quantum Electron.* 1987, 23, 123−129.

(32) Georgas, M.; Leu, J.; Moss, B.; Sun, C.; Stojanovic, V. Addressing Link-Level Design Tradeoffs for Integrated Photonic Interconnects. In *2011 IEEE Custom Integrated Circuits Conference (CICC)*; IEEE, 2011; pp 1−8.

(33) Ding, R.; Liu, Y.; Li, Q.; Xuan, Z.; Ma, Y.; Yang, Y.; Lim, A. E.-J.; Lo, G.-Q.; Bergman, K.; Baehr-Jones, T.; et al. A Compact Low-Power 320-Gb/s WDM Transmitter Based on Silicon Microrings. *IEEE Photonics J.* 2014, 6, 1−8.

(34) Wu, R.; Seyedi, M. A.; Wang, Y.; Hulme, J.; Fiorentino, M.; Beausoleil, R. G.; Cheng, K.-T. Pairing of Microring-Based Silicon Photonic Transceivers for Tuning Power Optimization. In *2018 23rd Asia and South Pacific Design Automation Conference (ASP-DAC)*; IEEE, 2018; Vol. 2018-Janua, pp 135−140.

(35) Takenaka, M.; Han, J.-H.; Boeuf, F.; Park, J.-K.; Li, Q.; Ho, C. P.; Lyu, D.; Ohno, S.; Fujikata, J.; Takahashi, S.; et al. III−V/Si Hybrid MOS Optical Phase Shifter for Si Photonic Integrated Circuits. *J. Lightwave Technol.* 2019, 37, 1474−1483.

(36) Miller, S.; Lee, Y.-H. D.; Cardenas, J.; Gaeta, A. L.; Lipson, M. Electro-Optic Effect in Silicon Nitride. In *CLEO: 2015*; OSA: Washington, D.C., 2015; p SF1G.4.

(37) Berciano, M.; Marcaud, G.; Damas, P.; Le Roux, X.; Crozat, P.; Alonso Ramos, C.; Pérez Galacho, D.; Benedikovic, D.; Marris-Morini, D.; Cassan, E.; et al. Fast Linear Electro-Optic Effect in a Centrosymmetric Semiconductor. *Commun. Phys.* 2018, 1, 64.

(38) Hermans, A.; Van Daele, M.; Dendooven, J.; Clemmen, S.; Detavernier, C.; Baets, R. Integrated Silicon Nitride Electro-Optic Modulators with Atomic Layer Deposited Overlays. *Opt. Lett.* 2019, 44, 1112.

(39) Abel, S.; Eltes, F.; Ortmann, J. E.; Messner, A.; Castera, P.; Wagner, T.; Urbonas, D.; Rosa, A.; Gutierrez, A. M.; Tulli, D.; et al. Large Pockels Effect in Micro- and Nanostructured Barium Titanate Integrated on Silicon. *Nat. Mater.* 2019, 18, 42−47.

(40) Meier, A. R.; Niu, F.; Wessels, B. W. Integration of $BaTiO_3$ on Si (001) Using MgO/STO Buffer Layers by Molecular Beam Epitaxy. *J. Cryst. Growth* 2006, 294, 401−406.

(41) Abel, S.; Stöferle, T.; Marchiori, C.; Rossel, C.; Rossell, M. D.; Erni, R.; Caimi, D.; Sousa, M.; Chelnokov, A.; Offrein, B. J.; et al. A Strong Electro-Optically Active Lead-Free Ferroelectric Integrated on Silicon. *Nat. Commun.* 2013, 4, 1671.

(42) Eltes, F.; Caimi, D.; Fallegger, F.; Sousa, M.; O'Connor, E.; Rossell, M. D.; Offrein, B.; Fompeyrine, J.; Abel, S. Low-Loss $BaTiO_3$–Si Waveguides for Nonlinear Integrated Photonics. *ACS Photonics* 2016, 3, 1698−1703.

(43) Pernice, W. H. P.; Xiong, C.; Walker, F. J.; Tang, H. X. Design of a Silicon Integrated Electro-Optic Modulator Using Ferroelectric $BaTiO_3$ Films. *IEEE Photonics Technol. Lett.* 2014, 26, 1344−1347.

(44) Abel, S.; Stoferle, T.; Marchiori, C.; Caimi, D.; Czornomaz, L.; Stuckelberger, M.; Sousa, M.; Offrein, B. J.; Fompeyrine, J. A Hybrid Barium Titanate−Silicon Photonics Platform for Ultraefficient Electro-Optic Tuning. *J. Lightwave Technol.* 2016, 34, 1688−1693.

(45) Eltes, F.; Kroh, M.; Caimi, D.; Mai, C.; Popoff, Y.; Winzer, G.; Petousi, D.; Lischke, S.; Ortmann, J. E.; Czornomaz, L.; et al. A Novel 25 Gbps Electro-Optic Pockels Modulator Integrated on an Advanced Si Photonic Platform. In *2017 IEEE International Electron Devices Meeting (IEDM)*; IEEE, 2017; pp 24.5.1−24.5.4.

(46) Eltes, F.; Mai, C.; Caimi, D.; Kroh, M.; Popoff, Y.; Winzer, G.; Petousi, D.; Lischke, S.; Ortmann, J. E.; Czornomaz, L.; et al. A $BaTiO_3$-Based Electro-Optic Pockels Modulator Monolithically Integrated on an Advanced Silicon Photonics Platform. *J. Lightwave Technol.* 2019, 37, 1456−1462.

(47) Alexander, K.; George, J. P.; Verbist, J.; Neyts, K.; Kuyken, B.; Van Thourhout, D.; Beeckman, J. Nanophotonic Pockels Modulators on a Silicon Nitride Platform. *Nat. Commun.* 2018, 9, 3444.

(48) Wang, C.; Zhang, M.; Chen, X.; Bertrand, M.; Shams-Ansari, A.; Chandrasekhar, S.; Winzer, P.; Lončar, M. Integrated Lithium Niobate Electro-Optic Modulators Operating at CMOS-Compatible Voltages. *Nature* 2018, 562, 101.

(49) Tang, P.; Towner, D. J.; Meier, A. L.; Wessels, B. W. Polarisation-Insensitive $Si_3N_4$ Strip-Loaded $BaTiO_3$ Thin-Film Waveguide with Low Propagation Losses. *Electron. Lett.* 2003, 39, 1651.

(50) Niu, F.; Teren, A. R.; Hoerman, B. H.; Wessels, B. W. Epitaxial Ferroelectric $BaTiO_3$ Thin Films for Microphotonic Applications. *MRS Proc.* 2000, 637, E1.9.

(51) Petraru, A.; Schubert, J.; Schmid, M.; Buchal, C. Ferroelectric $BaTiO_3$ Thin-Film Optical Waveguide Modulators. *Appl. Phys. Lett.* 2002, 81, 1375−1377.

(52) Petraru, A.; Schubert, J.; Schmid, M.; Trithaveesak, O.; Buchal, C. Integrated Optical Mach Zehnder Modulator Based on Polycrystalline $BaTiO_3$. *Opt. Lett.* 2003, 28, 2527.

(53) Wessels, B. W. Ferroelectric Epitaxial Thin Films for Integrated Optics. *Annu. Rev. Mater. Res.* 2007, 37, 659−679.

(54) Dong, P.; Qian, W.; Liang, H.; Shafiiha, R.; Feng, N.-N.; Feng, D.; Zheng, X.; Krishnamoorthy, A. V.; Asghari, M. Low Power and Compact Reconfigurable Multiplexing Devices Based on Silicon Microring Resonators. *Opt. Express* 2010, 18, 9852.

(55) Kormondy, K. J.; Popoff, Y.; Sousa, M.; Eltes, F.; Caimi, D.; Rossell, M. D.; Fiebig, M.; Hoffmann, P.; Marchiori, C.; Reinke, M.; et al. Microstructure and Ferroelectricity of $BaTiO_3$ Thin Films on Si for Integrated Photonics. *Nanotechnology* 2017, 28, 075706.

(56) Abel, S.; Sousa, M.; Rossel, C.; Caimi, D.; Rossell, M. D.; Erni, R.; Fompeyrine, J.; Marchiori, C. Controlling Tetragonality and Crystalline Orientation in $BaTiO_3$ Nano-Layers Grown on Si. *Nanotechnology* 2013, 24, 285701.

(57) Dubourdieu, C.; Bruley, J.; Arruda, T. M.; Posadas, A.; Jordan-Sweet, J.; Frank, M. M.; Cartier, E.; Frank, D. J.; Kalinin, S. V.; Demkov, A. A.; et al. Switching of Ferroelectric Polarization in Epitaxial $BaTiO_3$ Films on Silicon without a Conducting Bottom Electrode. *Nat. Nanotechnol.* 2013, 8, 748−754.

(58) Xu, Q.; Schmidt, B.; Pradhan, S.; Lipson, M. Micrometre-Scale Silicon Electro-Optic Modulator. *Nature* 2005, 435, 325−327.

(59) Chen, L.; Wood, M. G.; Reano, R. M. 12.5 Pm/V Hybrid Silicon and Lithium Niobate Optical Microring Resonator with Integrated Electrodes. *Opt. Express* 2013, 21, 27003.

(60) Chen, L.; Xu, Q.; Wood, M. G.; Reano, R. M. Hybrid Silicon and Lithium Niobate Electro-Optical Ring Modulator. *Optica* **2014**, *1*, 112.
(61) Ahmed, A. N. R.; Shi, S.; Zablocki, M.; Yao, P.; Prather, D. W. Tunable Hybrid Silicon Nitride and Thin-Film Lithium Niobate Electro-Optic Microresonator. *Opt. Lett.* **2019**, *44*, 618.
(62) Hsu, M.-H. M.; Marinelli, A.; Merckling, C.; Pantouvaki, M.; Van Campenhout, J.; Absil, P.; Van Thourhout, D. Orientation-Dependent Electro-Optical Response of $BaTiO_3$ on $SrTiO_3$-Buffered Si(001) Studied via Spectroscopic Ellipsometry. *Opt. Mater. Express* **2017**, *7*, 2030.
(63) Ieda, M.; Sawa, G.; Kato, S. A Consideration of Poole-Frenkel Effect on Electric Conduction in Insulators. *J. Appl. Phys.* **1971**, *42*, 3737−3740.
(64) Niu, G.; Gautier, B.; Yin, S.; Saint-Girons, G.; Lecoeur, P.; Pillard, V.; Hollinger, G.; Vilquin, B. Molecular Beam Epitaxy Growth of $BaTiO_3$ Thin Films and Crucial Impact of Oxygen Content Conditions on the Electrical Characteristics. *Thin Solid Films* **2012**, *520*, 4595−4599.
(65) Hamann, H. F.; Lacey, J.; Weger, A.; Wakil, J. Spatially-Resolved Imaging of Microprocessor Power (SIMP): Hotspots in Microprocessors. In *Thermal and Thermomechanical Proceedings 10th Intersociety Conference on Phenomena in Electronics Systems, 2006.* ITHERM 2006; IEEE, 2006; Vol. *2006*, pp 121−125.
(66) Cocorullo, G.; Della Corte, F. G.; Rendina, I. Temperature Dependence of the Thermo-Optic Coefficient in Crystalline Silicon between Room Temperature and 550 K at the Wavelength of 1523 Nm. *Appl. Phys. Lett.* **1999**, *74*, 3338−3340.
(67) Della Corte, F. G.; Esposito Montefusco, M.; Moretti, L.; Rendina, I.; Cocorullo, G. Temperature Dependence Analysis of the Thermo-Optic Effect in Silicon by Single and Double Oscillator Models. *J. Appl. Phys.* **2000**, *88*, 7115−7119.
(68) Komma, J.; Schwarz, C.; Hofmann, G.; Heinert, D.; Nawrodt, R. Thermo-Optic Coefficient of Silicon at 1550 Nm and Cryogenic Temperatures. *Appl. Phys. Lett.* **2012**, *101*, 041905.
(69) DeRose, C. T.; Martinez, N. J.; Kekatpure, R. D.; Zortman, W. A.; Starbuck, A. L.; Pomerene, A.; Lentine, A. L. Thermal Crosstalk Limits for Silicon Photonic DWDM Interconnects. In *2014 Optical Interconnects Conference*; IEEE, 2014; Vol. *5*, pp 125−126.
(70) Messner, A.; Eltes, F.; Ma, P.; Abel, S.; Baeuerle, B.; Josten, A.; Heni, W.; Caimi, D.; Fompeyrine, J.; Leuthold, J. Integrated Ferroelectric $BaTiO_3$/Si Plasmonic Modulator for 100 Gbit/s and Beyond. In *Optical Fiber Communication Conference*; OSA: Washington, D.C., 2018; Vol. Part F84-O, p M2I.6.
(71) Hryniewicz, J. V.; Absil, P. P.; Little, B. E.; Wilson, R. A.; Ho, P.-T. Higher Order Filter Response in Coupled Microring Resonators. *IEEE Photonics Technol. Lett.* **2000**, *12*, 320−322.
(72) Poon, J. K.; Zhu, L.; DeRose, G. a; Yariv, A. Transmission and Group Delay of Microring Coupled-Resonator Optical Waveguides. *Opt. Lett.* **2006**, *31*, 456.
(73) Xia, F.; Sekaric, L.; Vlasov, Y. Ultracompact Optical Buffers on a Silicon Chip. *Nat. Photonics* **2007**, *1*, 65−71.
(74) Xia, F.; Rooks, M.; Sekaric, L.; Vlasov, Y. Ultra-Compact High Order Ring Resonator Filters Using Submicron Silicon Photonic Wires for on-Chip Optical Interconnects. *Opt. Express* **2007**, *15*, 11934.
(75) Goykhman, I.; Desiatov, B.; Levy, U. Ultrathin Silicon Nitride Microring Resonator for Biophotonic Applications at 970 Nm Wavelength. *Appl. Phys. Lett.* **2010**, *97*, 081108.
(76) Heck, M. J. R. Highly Integrated Optical Phased Arrays: Photonic Integrated Circuits for Optical Beam Shaping and Beam Steering. *Nanophotonics* **2017**, *6*, 93−107.
(77) Poulton, C. V.; Yaacobi, A.; Cole, D. B.; Byrd, M. J.; Raval, M.; Vermeulen, D.; Watts, M. R. Coherent Solid-State LIDAR with Silicon Photonic Optical Phased Arrays. *Opt. Lett.* **2017**, *42*, 4091.